%% file: main.tex
\documentclass[sigplan,dvipsnames]{acmart}

\acmConference[PL'18]{ACM SIGPLAN Conference on Programming Languages}{January 01--03, 2018}{New York, NY, USA}
\acmYear{2018}
\acmISBN{} %
\acmDOI{} %
\startPage{1}

\setcopyright{none}

\renewcommand\footnotetextcopyrightpermission[1]{} %
\usepackage{booktabs}   %
\usepackage{subcaption} %
\usepackage{paralist}

\usepackage{listings}
\usepackage{enumitem}

\usepackage{todonotes}
\usepackage{bm}
\usepackage{cleveref}
\usepackage{balance}

\include{macros}

\newcommand{\src}[1]{\archStyle{#1}}
\newcommand{\trg}[1]{\muarchStyle{#1}}
\newcommand{\trgb}[1]{\ensuremath{\bm{\col{\ulccol }{#1}}}}
\newcommand{\con}[1]{\interfStyle{#1}}

\newcommand{\bl}[1]{\col{\neutcol }{#1}}

\newcommand{\archSemP}[1]{\archStyle{ATR\left(\bl{#1}\right)} }
\newcommand{\muarchSemP}[1]{\muarchStyle{HTR\left(\bl{#1}\right) } }
\newcommand{\interfSemP}[1]{\interfStyle{CTR_{c}\left(\bl{#1}\right) }}

\newcommand{\hs}[0]{\Sigma}
\newcommand{\cs}[0]{\Xi}

\renewcommand{\paragraph}[1]{\noindent~\textbf{#1:}~}

\newcommand{\CtSeqInterfP}[1]{\con{c^{seq}_{ct}}}
\newcommand{\CtSpecInterfP}[1]{\con{c^{spec}_{ct}}}
\newcommand{\ArchSeqInterfP}[1]{\con{c^{seq}_{arch}}}
\newcommand{\CtPcSpecInterfP}[1]{\con{c^{seq-spec}_{ct-pc}}}

\begin{document}

\title[]{Contract-Aware Secure Compilation}         %

\author{Marco Guarnieri}
\affiliation{
  \institution{IMDEA Software Institute}            %
}

\author{Marco Patrignani}
\affiliation{
  \institution{Stanford University}           %
\institution{CISPA Helmholz Center for Information Security}
}

\maketitle

\section{Introduction}

Microarchitectural attacks, such as Spectre~\cite{Kocher2018spectre}, Meltdown~\cite{meltdown2018}, Foreshadow~\cite{vanbulck2018foreshadow}, RIDL~\cite{ridl2019}, and ZombieLoad~\cite{zombieload2019}, exploit the abstraction gap between the Instruction Set Architecture (ISA) and how instructions are actually executed by processors to compromise the confidentiality and integrity of a system. 
That is, these attacks exploit microarchitectural side-effects resulting from processor's optimizations, such as speculative  and out-of-order execution, and from processor's internal buffers and caches that are invisible at the ISA level.

To secure systems against microarchitectural attacks,  programmers need to reason about and \textit{program against} these microarchitectural side-effects.
There is, however, no ``unique'' reference for microarchitectural side-effects.
Even for a single manufacturer, processors subtly differ in security-relevant microarchitectural side-effects across generations.
For example, the \texttt{clflush} instruction for flushing caches behaves differently across generations of Intel processors~\cite{flushgeist}.
As a result, a program might be secure when run on a  processor and insecure when run on another processor providing slightly different  guarantees.
However, we cannot---and should not---expect programmers to manually tailor programs for specific processors and their security guarantees.

Instead, we could rely on compilers (and the secure compilation community), as they can play a prominent role in bridging this gap:
compilers should target specific processors microarchitectural security guarantees and they should leverage these guarantees to produce secure code. %
This will enable decoupling program-level security (say, ensuring that secrets are not leaked under the ISA semantics), which programmers should enforce, and microarchitectural security (say, preventing leaks of secrets due to microarchitectural side-effects), which is the job of the compiler.

To achieve this, we outline the idea of \textit{Contract-Aware Secure COmpilation} (CASCO) where compilers are parametric with respect to a hardware/software security-contract, an abstraction capturing a processor's  security guarantees.
That is, compilers will automatically leverage the guarantees formalized in the contract to ensure that program-level security properties are preserved at microarchitectural level. 

For concreteness, our overview of CASCO builds on a recent formulation of hardware/software contracts~\cite{guarnieri2021contracts} that focuses on  data confidentiality (and therefore hypersafety properties).
We believe that  CASCO is more general and it can be applied  also to other classes of security properties.
 \section{Contract-aware secure compilation}
The CASCO framework relies on the following elements: ISA, Hardware and Contract languages, the adversary we consider and the notion of contract-aware compilers.

\smallskip
\paragraph{ISA language}
We consider an ISA language \src{L} with a notion of programs \src{p} (comprising both code and data segments) and of architectural program states $\src{\sigma} \in \src{AS}$.
\src{L} is equipped with an \textit{architectural semantics} $\archStepCompact:\src{AS}\times\src{AS}$ that models the execution of programs at the architectural level, mapping an \archstate{} $\src{\sigma}$ to its successor $\src{\sigma'}$. %
Assume given $\archSemP{\src{p}}$, a function that denotes the \emph{Architectural TRaces} of \src{p}, derived from the sequence of \archstate{}s $\src{\sigma} \cdot \ldots \cdot \src{\sigma_n}$ that the execution of \src{\sigma} goes through according to $\archStepCompact$.

\smallskip
\paragraph{Hardware}
The execution of $\src{L}$-programs at the microarchitectural level is formalised with a \textit{hardware semantics} that relies on {\em hardware states} $\trgb{\hs}=\trg{\tup{\src{\sigma},\trgb{\mu}}} \in \trg{HS}$.
Hardware states consist of an \archstate{} $\src{\sigma}$ (as before) and a {\em microarchitectural state} $\trgb{\mu}$, which models the state of components like predictors, caches, and  reorder buffers. 
A hardware semantics $\muarchStepCompact{} : \trg{HS} \times \trg{HS}$ 
maps
hardware states $\trgb{\hs}$  to their successor $\trgb{\hs'}$.\looseness=-1 %

\smallskip
\paragraph{Adversary}
We consider a hardware-level adversary that can observe parts of the \uarchstate{} during execution.
Given a program $\src{p}$, 
$\muarchSemP{\src{p}}$ denotes the \emph{Hardware TRaces} of \src{p}, that is, the sequence of hardware observations
$\adversary(\trg{\trgb{\mu}_0}) \concat \ldots \concat \adversary(\trg{\trgb{\mu}_n})$
that the hardware state \trgb{\hs} of \src{p} goes through according to \muarchStepCompact{}. 
Here, $\adversary(\trgb{\mu})$ maps $\trgb{\mu}$ to its attacker-visible components (say, the cache metadata). %

\smallskip
\paragraph{Contracts}
A contract splits the responsibilities for preventing side-channels between software and hardware, and it provides a concise representation of a processor's microarchitectural security guarantees.
Following~\cite{guarnieri2021contracts}, a {\em contract} \con{c} defines:
\begin{inparaenum}[(1)]
\item a notion of contract states $\con{\cs}\in\con{CS}$ that extend \src{\sigma} with contract-related components,
\item labels $\con{l}\in\con{LC}$ representing contract-observations, and 
\item  a labeled semantics $\interfStepCompact : \con{CS}\times\con{LC}\times\con{CS}$.
\end{inparaenum}
Given a program $\src{p}$, \interfSemP{\src{p}} denotes the \emph{Contract TRace} of \src{p}, that is, the sequence \con{l_1,\cdots,l_n} of labels that the contract state \con{\cs} of \src{p} goes through according to \interfStepCompact{}. 

The contract traces of a program $\interfSemP{\cdot}$ capture which \archstate{}s are guaranteed to be indistinguishable by a hardware attacker on any hardware platform satisfying the contract: %
\begin{definition}[Hardware satisfies contract~\cite{guarnieri2021contracts}]\label{def:hni}
A hardware semantics $\muarchSemP{\cdot}$ {\em satisfies a contract  $\con{c}$} (denoted $\hsni{\con{c}}{\muarchSemP{\cdot}}$) if, for all programs \src{p} and \src{p'} that only vary in the data segment, if $\interfSemP{\src{p}} = \interfSemP{\src{p'}}$, then $\muarchSemP{\src{p}}= \muarchSemP{\src{p'}}$.
\end{definition}

\paragraph{Contract-aware compilers}
Contract-aware compilers~($\comp{\cdot}$) are parametric with respect to a contract $\con{c}\in\con{\mk{C}}$, which formalizes a processor's security guarantees. %
The target program \comp{\con{c},\src{p}} depends on the source $\src{p}$ and on the contract $\con{c}$.\looseness=-1

Contract-aware compilers can be constructed to preserve many security properties (e.g., cryptographic constant-time and absence of speculative leaks), so long as these properties are expressible in the contract semantics (fortunately, this is often the case~\cite{guarnieri2021contracts}).
Depending on the property of interest, 
we then choose different secure compilation criteria and instantiate them with the ISA and contract semantics.
Proving that a contract-aware compiler upholds such a criterion demonstrates that the criterion is preserved for all contracts in $\con{c} \in \con{\mk{C}}$, which determine the target language's semantics.

As an example, consider the security property of interest being the prevention of all microarchitectural leaks of information not exposed by ISA observations (captured by the architectural traces $\archSemP{\cdot}$); this can ensure, for instance, the absence of leaks of transiently accessed data~\cite{patrignani2019exorcising}.
We therefore choose the secure compilation criterion preserving 2-hypersafety properties~\cite{rhc,rhc-rel}.
An instantiation of that criterion is found \Cref{def:sec-comp} below.
That informally tells that the compiler translates ISA-equivalent programs into contract-equivalent ones, so there is no more leakage at the contract level than what expressable in the ISA.\looseness=-1

\begin{definition}[Compiler satisfies contract]\label{def:sec-comp}
We say that a compiler $\comp{\cdot}$ is secure for all contracts of $\con{\mk{C}}$
(denoted as $\comp{\cdot} \vdash {\con{\mk{C}}}$)
if for all contracts $\con{c} \in \con{\mk{C}}$ and programs $\src{p}$, \src{p'} that only differ in the data segment, 
if $\archSemP{\src{p}} = \archSemP{\src{p'}}$, 
then $\interfSemP{\comp{\con{c},\src{p}}} = \interfSemP{\comp{\con{c},\src{p'}}}$.
\end{definition}

\Cref{theorem:main} illustrates the overarching benefits of using CASCO.
It is sufficient to show  that both the hardware and the compiler satisfy a contract (\Cref{def:hni} and \Cref{def:sec-comp}) to derive that any ISA program \src{p} will produce hardware executions that will not be vulnerable to attacks when run on hardware satisfying the contract.
Notably, proofs of \Cref{def:hni} and \Cref{def:sec-comp} can be done separately and by different parties:
hardware developers can provide contracts and proving \Cref{def:hni} independently of specific compiler criteria, while developers of secure compilers can focus on proving \Cref{def:sec-comp} ignoring most of the hardware details (except those captured by contracts).
\begin{theorem}\label{theorem:main}
	If  
    $\comp{\cdot} \vdash {\con{\mk{C}}}$, $\con{c}\in\con{\mk{C}}$, and $\hsni{\con{c}}{\muarchSemP{\cdot}}$,
    then for all programs $\src{p}$ and \src{p'} that only differ in the data segment, 
    if $\archSemP{\src{p}} = \archSemP{\src{p'}}$, 
    then $\muarchSemP{\comp{\con{c},\src{p}}} = \muarchSemP{\comp{\con{c}, \src{p'}}}$.
\end{theorem}

\section{CASCO for secure speculation}

To illustrate the benefits of CASCO, we focus on speculative execution attacks (Spectre) as an example due to the availability of compiler-level countermeasures~\cite{patrignani2019exorcising} and security contracts~\cite{guarnieri2021contracts}.
CASCO, however, is more general and it can be applied to all settings where microarchitectural attacks are prevented by compiler-inserted countermeasures.

Consider the classical Spectre v1 attack~\cite{Kocher2018spectre}.
There, an \trg{attacker} poisons the \trg{branch} \trg{predictor} (which exists at the \trg{hardware} level and not at the \src{ISA}) to trigger speculative execution and encode speculatively accessed data (otherwise unaccessible) into the \trg{cache}, so the attacker can later retrieve them by probing the \trg{cache}.

There exist four different \con{contracts} that serve as specifications of processors' microarchitectural security guarantees~\cite{guarnieri2021contracts} and that compilers can use as security specification.\looseness=-1

\begin{description}[leftmargin=0pt, parsep=0pt, listparindent=1em, font =\sffamily\bfseries]

\item[Contract $\CtSeqInterfP{\cdot}$:]
This contract exposes the program counter and the locations of memory accesses on sequential, non-speculative paths.
$\CtSeqInterfP{\cdot}$ is often used to formalize constant-time programming~\cite{BartheBCL19,AlmeidaBBDE16}, and it is satisfied (in the sense of \Cref{def:hni}) by in-order, non-speculative processors~\cite{guarnieri2021contracts}.

\item[Contract $\CtSpecInterfP{\cdot}$:]
	This contract additionally exposes the program counter and the locations of all memory accesses on speculatively executed paths~\cite{guarnieri2020spectector}. Simple speculative out-of-order processors satisfy $\CtSpecInterfP{\cdot}$~\cite{guarnieri2021contracts}.

\item[Contract $\ArchSeqInterfP{\cdot}$:]
This contract, which guarantees the confidentiality of data that is {\em only transiently} loaded,  exposes the program counter, the location of all loads and stores, and the values of all data loaded from memory on standard, i.e., non-speculative, program paths.
Processors implementing speculative taint tracking~\cite{STT2019,nda2019weisse} satisfy $\ArchSeqInterfP{\cdot}$~\cite{guarnieri2021contracts}.

\item[Contract $\CtPcSpecInterfP{\cdot}$:]
This contract exposes program counter and addresses of loads during sequential execution, and only the program counter during speculative execution.
Processors with load-delay countermeasures~\cite{specshadow2019} satisfy $\CtPcSpecInterfP{\cdot}$~\cite{guarnieri2021contracts}.
\end{description}

A possible countermeasure against Spectre v1 attacks, implemented in the Microsoft Visual C++ and Intel ICC compilers~\cite{Intel-compiler,microsoft}, is the insertion of \texttt{lfence} instructions (which stop speculation).
The countermeasure has been developed to work against speculative, out-of-order processors (contract $\CtSpecInterfP{\cdot}$), and it  injects an \texttt{lfence} instruction after all branch instructions, preventing the attack described before.
However, a contract-aware compiler can rely on the contract information to know the underlying processor's security guarantees and optimise its code, avoiding the injection of unnecessary \texttt{lfence}s. 
For example, consider processors that implement load-delay (contract $\CtPcSpecInterfP{\cdot}$) or speculative taint-tracking countermeasures (contract $\ArchSeqInterfP{\cdot}$). 
A contract-aware compiler targeting those processors can avoid inserting \texttt{lfence}s after branches since there, the speculative memory leaks are prevented by the hardware.

\section{Future directions}

We believe CASCO provides foundations for designing and proving the correctness of compilers that automatically leverage hardware-level security guarantees, formalized using security contracts, to prevent microarchitectural leaks. %
For this, we will need
\begin{inparaenum}[(1)]
\item formal languages for modeling interesting classes of contracts;
\item ways of formalizing compilers that use contract information to optimise code; and %
\item new proof techniques that account for contract parametricity and composability (to simplify proofs across similar contracts).
\end{inparaenum}

\smallskip
{\small
\textbf{Acknowledgements:}
This work was partially supported by the German Federal Ministry of Education and Research (BMBF) through funding for the CISPA-Stanford Center for Cybersecurity (FKZ: 13N1S0762), by a grant from Intel Corporation, Juan de la Cierva-Formaci\'on grant FJC2018-036513-I, Spanish project RTI2018-102043-B-I00 SCUM, and Madrid regional project S2018/TCS-4339 BLOQUES.
}

\balance
\bibliography{bibfile}

\end{document}

%% file: macros.tex
\definecolor{ao(english)}{rgb}{0.0, 0.5, 0.0}
\definecolor{royalblue(web)}{rgb}{0.25, 0.41, 0.88}

\definecolor{codegray}{gray}{0.95}

\theoremstyle{definition}
\newtheorem{definition}{Definition}

\theoremstyle{theorem}
\newtheorem{theorem}{Theorem}

\newcommand{\tup}[1]{\langle #1 \rangle}

\newcommand{\concat}{\cdot}

\renewcommand{\paragraph}[1]{\smallskip\noindent\textbf{#1:\ }}

\newcommand{\kywd}[1]{\mathbf{#1}}

\definecolor{Blue3}{HTML}{0000CD}
\newcommand{\obsKywd}[1]{\textcolor{Blue3}{\mathtt{#1}}}

\newcommand{\startObsKywd}[1]{\obsKywd{start}}
\newcommand{\commitObsKywd}[1]{\obsKywd{commit}} 
\newcommand{\rollbackObsKywd}[1]{\obsKywd{rollback}}

\newcommand{\commitObs}[1]{\commitObsKywd{}} %
\newcommand{\rollbackObs}[1]{\rollbackObsKywd{}} %

\newcommand{\fetch}[1]{
\ifthenelse{\equal{#1}{}}{\kywd{fetch}}{\kywd{fetch}}%
}

\newcommand{\mi}[1]{\ensuremath{\mathit{#1}}}

\newcommand{\mf}[1]{\ensuremath{\mathbf{#1}}}
\newcommand{\mk}[1]{\ensuremath{\mathfrak{#1}}}

\newcommand{\ms}[1]{\ensuremath{\mathsf{#1}}}

\newcommand{\neutcol}[0]{black}
\newcommand{\stlccol}[0]{RoyalBlue}
\newcommand{\ulccol}[0]{RedOrange}

\newcommand{\idecol}[0]{Emerald}%

\newcommand{\col}[2]{\ensuremath{{\color{#1}{#2}}}}

\newcommand{\archStyle}[1]{\ms{\col{\stlccol}{#1}}}
\newcommand{\muarchStyle}[1]{{\mf{\col{\ulccol }{#1}}}}
\newcommand{\interfStyle}[1]{{\mi{\col{\idecol }{#1}}}}

\newcommand{\archStepCompact}{\archStyle{\rightarrow}}

\newcommand{\muarchStepCompact}{\muarchStyle{\Rightarrow}}

\newcommand{\interfStepCompact}{\interfStyle{\rightharpoonup}}

\newcommand{\hsni}[2]{#2 \vdash #1}

\newcommand{\adversary}{\mathcal{A}}

\definecolor{Blue3}{HTML}{0000CD}
\definecolor{Green4}{HTML}{008B00}
\definecolor{Red3}{HTML}{CD0000}
\definecolor{orange}{rgb}{0.8, 0.47, 0.196}

\usepackage{letltxmacro}
\newcommand*{\SavedLstInline}{}
\LetLtxMacro\SavedLstInline\lstinline
\DeclareRobustCommand*{\lstinline}{%
  \ifmmode
    \let\SavedBGroup\bgroup
    \def\bgroup{%
      \let\bgroup\SavedBGroup
      \hbox\bgroup
    }%
  \fi
  \SavedLstInline
}

\newcommand{\archstate}{architectural state}
\newcommand{\uarchstate}{microarchitectural state}

\newcommand{\techReportAppendix}[1]{%
	\ifthenelse{\equal{\shortVersion}{true}}%
		{\cite{technicalReport}}%
        {Appendix~\ref{#1}}%
}%

\newcommand{\techReportAppendices}[2]{%
	\ifthenelse{\equal{\shortVersion}{true}}%
		{\cite{technicalReport}}%
		{Appendices~\ref{#1}--\ref{#2}}%
}%

\newcommand{\onlyTechReport}[1]{%
	\ifthenelse{\equal{\shortVersion}{true}}%
		{}%
		{#1}%
}%

\newcommand{\onlyShortVersion}[1]{%
	\ifthenelse{\equal{\shortVersion}{true}}%
		{#1}%
		{}%
}%

\newcommand{\compskel}[3]{\ensuremath{\bl{\left(\!\left| \src{#1} \right|\!\right)^{#2}_{#3}}}}
\newcommand{\comp}[1]{\compskel{\bl{#1}}{}{}}

\definecolor{mygreen}{rgb}{0,0.6,0}
\definecolor{mygray}{rgb}{0.5,0.5,0.5}
\definecolor{mymauve}{rgb}{0.58,0,0.82}

\lstdefinelanguage{Java} %
{morekeywords={abstract, all, and, as, assert, but, disj, else, exactly, extends, fact, for, fun, iden, if, iff, implies, in, Int, void, int, let, lone, module, no, none, not, one, open, or, part, pred, run, seq, set, sig, some, sum, then, univ, package, class, public, private, null, return, new, interface, extern, object, implements, System, static, super, try , catch, throw, throws, Unit, var, val, of, principal, trust},
sensitive=true,
keywordstyle=\bfseries\color{\stlccol}, %
commentstyle=\itshape\color{purple!40!black},
morecomment=[l][\small\itshape\color{purple!40!black}]{//},
identifierstyle=\color{\stlccol},
stringstyle=\color{orange},
basicstyle=\small,
basicstyle={\small\ttfamily},
numbers=left,
numberstyle=\tiny\color{mygray},
tabsize=2,
numbersep=5pt,
breaklines=true,
lineskip=-2pt,
stepnumber=1,
captionpos=b,
breaklines=true,
breakatwhitespace=false,
showspaces=false,
showtabs=false,
float=!h,
columns=fullflexible,escapeinside={(*@}{@*)},
moredelim=**[is][\color{red!60}]{@}{@},
literate={->}{{$\to$}}1 {^}{{$\mspace{-3mu}\widehat{\quad}\mspace{-3mu}$}}1
{<}{$<$ }2 {>}{$>$ }2 {>=}{$\geq$ }2 {=<}{$\leq$ }2
{<:}{{$<\mspace{-3mu}:$}}2 {:>}{{$:\mspace{-3mu}>$}}2
{=>}{{$\Rightarrow$ }}2 {+}{$+$ }2 {++}{{$+\mspace{-8mu}+$ }}2
{<=>}{{$\Leftrightarrow$ }}2 {+}{$+$ }2 {++}{{$+\mspace{-8mu}+$ }}2
{\~}{{$\mspace{-3mu}\widetilde{\quad}\mspace{-3mu}$}}1
{!=}{$\neq$ }2 {*}{${}^{\ast}$}1 %
{\#}{$\#$}1
}

\lstdefinelanguage{General} %
{morekeywords={abstract, all, and, as, assert, but, disj, else, exactly, extends, fact, for, fun, iden, if, iff, implies, in, Int, void, int, let, lone, module, no, none, not, one, open, or, part, pred, run, seq, set, sig, some, sum, then, univ, package, class, public, private, null, return, new, interface, extern, object, implements, System, static, super, try , catch, throw, throws, Unit, var, val, of, principal, trust},
sensitive=true,
keywordstyle=\bfseries\color{\neutcol},
commentstyle=\itshape\color{purple!40!black},
morecomment=[l][\small\itshape\color{purple!40!black}]{//},
identifierstyle=\color{\neutcol},
stringstyle=\color{orange},
basicstyle=\small,
basicstyle={\small\ttfamily},
numbers=left,
numberstyle=\tiny\color{mygray},
tabsize=2,
numbersep=5pt,
breaklines=true,
lineskip=-2pt,
stepnumber=1,
captionpos=b,
breaklines=true,
breakatwhitespace=false,
showspaces=false,
showtabs=false,
float=!h,
columns=fullflexible,escapeinside={(*@}{@*)},
moredelim=**[is][\color{red!60}]{@}{@},
literate={->}{{$\to$}}1 {^}{{$\mspace{-3mu}\widehat{\quad}\mspace{-3mu}$}}1
{<}{$<$ }2 {>}{$>$ }2 {>=}{$\geq$ }2 {=<}{$\leq$ }2
{<:}{{$<\mspace{-3mu}:$}}2 {:>}{{$:\mspace{-3mu}>$}}2
{=>}{{$\Rightarrow$ }}2 {+}{$+$ }2 {++}{{$+\mspace{-8mu}+$ }}2
{<=>}{{$\Leftrightarrow$ }}2 {+}{$+$ }2 {++}{{$+\mspace{-8mu}+$ }}2
{\~}{{$\mspace{-3mu}\widetilde{\quad}\mspace{-3mu}$}}1
{!=}{$\neq$ }2 {*}{${}^{\ast}$}1 %
{\#}{$\#$}1
}

\lstdefinelanguage{Asm}
{morekeywords={abstract, all, and, as, assert, but, check, disj, else, exactly, extends, fact, for, fun, iden, if, iff, implies, in, Int, void, int, let, lone, module, no, none, not, one, open, part, pred, run, seq, set, sig, some, sum, then, univ, package, class, public, private, null, return, new, interface, extern, object, implements, System, static, super, try , catch, throw, throws, Unit, var, val, principal, trust, label, load, add, addi, into, test, mov, cmov, cmova, movzx, cmp, jbe, sar, cmovbe, or, jmp, shl, ret, jae, lea, lfence, jne},
sensitive=true,
identifierstyle=\color{\ulccol},
keywordstyle=\bfseries\color{\ulccol},
commentstyle=\itshape\color{purple!40!black},
morecomment=[l][\small\itshape\color{purple!40!black}]{//},
stringstyle=\color{orange},
basicstyle=\small,
basicstyle={\small},
numbers=left,
numberstyle=\tiny\color{mygray},
tabsize=2,
numbersep=5pt,
breaklines=true,
lineskip=-2pt,
stepnumber=1,
captionpos=b,
breaklines=true,
breakatwhitespace=false,
showspaces=false,
showtabs=false,
float=!h,
columns=fullflexible,escapeinside={(*@}{@*)},
moredelim=**[is][\color{red!60}]{@}{@},
literate={->}{{$\to$}}1 {^}{{$\mspace{-3mu}\widehat{\quad}\mspace{-3mu}$}}1
{<}{$<$ }2 {>}{$>$ }2 {>=}{$\geq$ }2 {=<}{$\leq$ }2
{<:}{{$<\mspace{-3mu}:$}}2 {:>}{{$:\mspace{-3mu}>$}}2
{=>}{{$\Rightarrow$ }}2 {+}{$+$ }2 {++}{{$+\mspace{-8mu}+$ }}2
{<=>}{{$\Leftrightarrow$ }}2 {+}{$+$ }2 {++}{{$+\mspace{-8mu}+$ }}2
{\~}{{$\mspace{-3mu}\widetilde{\quad}\mspace{-3mu}$}}1
{!=}{$\neq$ }2 {*}{${}^{\ast}$}1 %
{\#}{$\#$}1
}

%% file: main.bbl

\begin{thebibliography}{18}

%%% the Web address.

\ifx \showCODEN    \undefined \def \showCODEN     #1{\unskip}     \fi
\ifx \showDOI      \undefined \def \showDOI       #1{#1}\fi
\ifx \showISBNx    \undefined \def \showISBNx     #1{\unskip}     \fi
\ifx \showISBNxiii \undefined \def \showISBNxiii  #1{\unskip}     \fi
\ifx \showISSN     \undefined \def \showISSN      #1{\unskip}     \fi
\ifx \showLCCN     \undefined \def \showLCCN      #1{\unskip}     \fi
\ifx \shownote     \undefined \def \shownote      #1{#1}          \fi
\ifx \showarticletitle \undefined \def \showarticletitle #1{#1}   \fi
\ifx \showURL      \undefined \def \showURL       {\relax}        \fi
% invisible to TeX
\providecommand\bibfield[2]{#2}
\providecommand\bibinfo[2]{#2}
\providecommand\natexlab[1]{#1}
\providecommand\showeprint[2][]{arXiv:#2}

\bibitem[\protect\citeauthoryear{Abate, Blanco, Ciobaca, Durier, Garg,
  Hri\c{t}cu, Patrignani, , Tanter, and Thibault}{Abate et~al\mbox{.}}{2020}]%
        {rhc-rel}
\bibfield{author}{\bibinfo{person}{Carmine Abate}, \bibinfo{person}{Roberto
  Blanco}, \bibinfo{person}{Stefan Ciobaca}, \bibinfo{person}{Alexandre
  Durier}, \bibinfo{person}{Deepak Garg}, \bibinfo{person}{C\u{a}t\u{a}lin
  Hri\c{t}cu}, \bibinfo{person}{Marco Patrignani}, \bibinfo{person}{},
  \bibinfo{person}{Eric Tanter}, {and} \bibinfo{person}{J\'er\'emy Thibault}.}
  \bibinfo{year}{2020}\natexlab{}.
\newblock \showarticletitle{Trace-Relating Compiler Correctness and Secure
  Compilation}. In \bibinfo{booktitle}{\emph{ESOP 2020}}.
\newblock


\bibitem[\protect\citeauthoryear{Abate, Blanco, Garg, Hri\c{t}cu, Patrignani,
  and Thibault}{Abate et~al\mbox{.}}{2019}]%
        {rhc}
\bibfield{author}{\bibinfo{person}{Carmine Abate}, \bibinfo{person}{Roberto
  Blanco}, \bibinfo{person}{Deepak Garg}, \bibinfo{person}{C\u{a}t\u{a}lin
  Hri\c{t}cu}, \bibinfo{person}{Marco Patrignani}, {and}
  \bibinfo{person}{J\'er\'emy Thibault}.} \bibinfo{year}{2019}\natexlab{}.
\newblock \showarticletitle{Journey Beyond Full Abstraction: Exploring Robust
  Property Preservation for Secure Compilation}. In
  \bibinfo{booktitle}{\emph{CSF 2019}}.
\newblock


\bibitem[\protect\citeauthoryear{Almeida, Barbosa, Barthe, Dupressoir, and
  Emmi}{Almeida et~al\mbox{.}}{2016}]%
        {AlmeidaBBDE16}
\bibfield{author}{\bibinfo{person}{Jos{\'{e}}~Bacelar Almeida},
  \bibinfo{person}{Manuel Barbosa}, \bibinfo{person}{Gilles Barthe},
  \bibinfo{person}{Fran{\c{c}}ois Dupressoir}, {and} \bibinfo{person}{Michael
  Emmi}.} \bibinfo{year}{2016}\natexlab{}.
\newblock \showarticletitle{Verifying Constant-Time Implementations}. In
  \bibinfo{booktitle}{\emph{Proceedings of the 26th {USENIX} Security
  Symposium}} \emph{(\bibinfo{series}{{USENIX} Security'16})}.
  \bibinfo{publisher}{{USENIX} Association}.
\newblock


\bibitem[\protect\citeauthoryear{Barthe, Betarte, Campo, and Luna}{Barthe
  et~al\mbox{.}}{2019}]%
        {BartheBCL19}
\bibfield{author}{\bibinfo{person}{Gilles Barthe}, \bibinfo{person}{Gustavo
  Betarte}, \bibinfo{person}{Juan~Diego Campo}, {and} \bibinfo{person}{Carlos
  Luna}.} \bibinfo{year}{2019}\natexlab{}.
\newblock \showarticletitle{System-Level Non-interference of Constant-Time
  Cryptography. Part {I:} Model}.
\newblock \bibinfo{journal}{\emph{Journal of Automatic Reasoning}}
  \bibinfo{volume}{63}, \bibinfo{number}{1} (\bibinfo{year}{2019}).
\newblock


\bibitem[\protect\citeauthoryear{Guarnieri, K{\"o}pf, Morales, Reineke, and
  S\'{a}nchez}{Guarnieri et~al\mbox{.}}{2020a}]%
        {guarnieri2020spectector}
\bibfield{author}{\bibinfo{person}{Marco Guarnieri}, \bibinfo{person}{Boris
  K{\"o}pf}, \bibinfo{person}{Jos\'{e}~F. Morales}, \bibinfo{person}{Jan
  Reineke}, {and} \bibinfo{person}{Andr\'{e}s S\'{a}nchez}.}
  \bibinfo{year}{2020}\natexlab{a}.
\newblock \showarticletitle{\textsc{Spectector}: Principled detection of
  speculative information flows}. In \bibinfo{booktitle}{\emph{Proceedings of
  the 41st IEEE Symposium on Security and Privacy}}
  \emph{(\bibinfo{series}{S\&P'20})}. \bibinfo{publisher}{IEEE}.
\newblock


\bibitem[\protect\citeauthoryear{Guarnieri, K{\"o}pf, Reineke, and
  Vila}{Guarnieri et~al\mbox{.}}{2020b}]%
        {guarnieri2021contracts}
\bibfield{author}{\bibinfo{person}{Marco Guarnieri}, \bibinfo{person}{Boris
  K{\"o}pf}, \bibinfo{person}{Jan Reineke}, {and} \bibinfo{person}{Pepe Vila}.}
  \bibinfo{year}{2020}\natexlab{b}.
\newblock \showarticletitle{Hardware/Software Contracts for Secure
  Speculation}. In \bibinfo{booktitle}{\emph{Proceedings of the 42nd IEEE
  Symposium on Security and Privacy}} \emph{(\bibinfo{series}{S\&P'21})}.
  \bibinfo{publisher}{IEEE}.
\newblock


\bibitem[\protect\citeauthoryear{Intel}{Intel}{2018}]%
        {Intel-compiler}
\bibfield{author}{\bibinfo{person}{Intel}.} \bibinfo{year}{2018}\natexlab{}.
\newblock \bibinfo{title}{{Using Intel Compilers to Mitigate Speculative
  Execution Side-Channel Issues}}.
\newblock
  \bibinfo{howpublished}{https://software.intel.com/en-us/articles/using-intel-compilers-to-mitigate-speculative-execution-side-channel-issues}.
\newblock


\bibitem[\protect\citeauthoryear{Kocher, Horn, Fogh, Genkin, Gruss, Haas,
  Hamburg, Lipp, Mangard, Prescher, Schwarz, and Yarom}{Kocher
  et~al\mbox{.}}{2019}]%
        {Kocher2018spectre}
\bibfield{author}{\bibinfo{person}{Paul Kocher}, \bibinfo{person}{Jann Horn},
  \bibinfo{person}{Anders Fogh}, \bibinfo{person}{Daniel Genkin},
  \bibinfo{person}{Daniel Gruss}, \bibinfo{person}{Werner Haas},
  \bibinfo{person}{Mike Hamburg}, \bibinfo{person}{Moritz Lipp},
  \bibinfo{person}{Stefan Mangard}, \bibinfo{person}{Thomas Prescher},
  \bibinfo{person}{Michael Schwarz}, {and} \bibinfo{person}{Yuval Yarom}.}
  \bibinfo{year}{2019}\natexlab{}.
\newblock \showarticletitle{{Spectre Attacks: Exploiting Speculative
  Execution}}. In \bibinfo{booktitle}{\emph{Proceedings of the 40th IEEE
  Symposium on Security and Privacy}} \emph{(\bibinfo{series}{S\&P'19})}.
  \bibinfo{publisher}{IEEE}.
\newblock


\bibitem[\protect\citeauthoryear{Lipp, Schwarz, Gruss, Prescher, Haas, Fogh,
  Horn, Mangard, Kocher, Genkin, Yarom, and Hamburg}{Lipp
  et~al\mbox{.}}{2018}]%
        {meltdown2018}
\bibfield{author}{\bibinfo{person}{Moritz Lipp}, \bibinfo{person}{Michael
  Schwarz}, \bibinfo{person}{Daniel Gruss}, \bibinfo{person}{Thomas Prescher},
  \bibinfo{person}{Werner Haas}, \bibinfo{person}{Anders Fogh},
  \bibinfo{person}{Jann Horn}, \bibinfo{person}{Stefan Mangard},
  \bibinfo{person}{Paul Kocher}, \bibinfo{person}{Daniel Genkin},
  \bibinfo{person}{Yuval Yarom}, {and} \bibinfo{person}{Mike Hamburg}.}
  \bibinfo{year}{2018}\natexlab{}.
\newblock \showarticletitle{Meltdown: Reading Kernel Memory from User Space}.
  In \bibinfo{booktitle}{\emph{Proceedings of the 27th {USENIX} Security
  Symposium}} \emph{(\bibinfo{series}{{USENIX} Security'18})}.
  \bibinfo{publisher}{{USENIX} Association}.
\newblock


\bibitem[\protect\citeauthoryear{Pardoe}{Pardoe}{2018}]%
        {microsoft}
\bibfield{author}{\bibinfo{person}{Andrew Pardoe}.}
  \bibinfo{year}{2018}\natexlab{}.
\newblock \bibinfo{title}{Spectre mitigations in {MSVC}}.
\newblock
  \bibinfo{howpublished}{https://blogs.msdn.microsoft.com/vcblog/2018/01/15/spectre-mitigations-in-msvc/}.
\newblock


\bibitem[\protect\citeauthoryear{Patrignani and Guarnieri}{Patrignani and
  Guarnieri}{2019}]%
        {patrignani2019exorcising}
\bibfield{author}{\bibinfo{person}{Marco Patrignani} {and}
  \bibinfo{person}{Marco Guarnieri}.} \bibinfo{year}{2019}\natexlab{}.
\newblock \showarticletitle{Exorcising Spectres with Secure Compilers}.
\newblock \bibinfo{journal}{\emph{CoRR}}  \bibinfo{volume}{abs/1910.08607}
  (\bibinfo{year}{2019}).
\newblock


\bibitem[\protect\citeauthoryear{Sakalis, Kaxiras, Ros, Jimborean, and
  Sj\"{a}lander}{Sakalis et~al\mbox{.}}{2019}]%
        {specshadow2019}
\bibfield{author}{\bibinfo{person}{Christos Sakalis}, \bibinfo{person}{Stefanos
  Kaxiras}, \bibinfo{person}{Alberto Ros}, \bibinfo{person}{Alexandra
  Jimborean}, {and} \bibinfo{person}{Magnus Sj\"{a}lander}.}
  \bibinfo{year}{2019}\natexlab{}.
\newblock \showarticletitle{Efficient Invisible Speculative Execution Through
  Selective Delay and Value Prediction}. In
  \bibinfo{booktitle}{\emph{Proceedings of the 46th International Symposium on
  Computer Architecture}} \emph{(\bibinfo{series}{ISCA'19})}.
  \bibinfo{publisher}{ACM}.
\newblock


\bibitem[\protect\citeauthoryear{Schwarz, Lipp, Moghimi, Van~Bulck, Stecklina,
  Prescher, and Gruss}{Schwarz et~al\mbox{.}}{2019}]%
        {zombieload2019}
\bibfield{author}{\bibinfo{person}{Michael Schwarz}, \bibinfo{person}{Moritz
  Lipp}, \bibinfo{person}{Daniel Moghimi}, \bibinfo{person}{Jo Van~Bulck},
  \bibinfo{person}{Julian Stecklina}, \bibinfo{person}{Thomas Prescher}, {and}
  \bibinfo{person}{Daniel Gruss}.} \bibinfo{year}{2019}\natexlab{}.
\newblock \showarticletitle{{ZombieLoad}: Cross-Privilege-Boundary Data
  Sampling}. In \bibinfo{booktitle}{\emph{Proceedings of the 26th ACM SIGSAC
  Conference on Computer and Communications Security}}
  \emph{(\bibinfo{series}{CCS'19})}. \bibinfo{publisher}{ACM}.
\newblock


\bibitem[\protect\citeauthoryear{Van~Bulck, Minkin, Weisse, Genkin, Kasikci,
  Piessens, Silberstein, Wenisch, Yarom, and Strackx}{Van~Bulck
  et~al\mbox{.}}{2018}]%
        {vanbulck2018foreshadow}
\bibfield{author}{\bibinfo{person}{Jo Van~Bulck}, \bibinfo{person}{Marina
  Minkin}, \bibinfo{person}{Ofir Weisse}, \bibinfo{person}{Daniel Genkin},
  \bibinfo{person}{Baris Kasikci}, \bibinfo{person}{Frank Piessens},
  \bibinfo{person}{Mark Silberstein}, \bibinfo{person}{Thomas~F. Wenisch},
  \bibinfo{person}{Yuval Yarom}, {and} \bibinfo{person}{Raoul Strackx}.}
  \bibinfo{year}{2018}\natexlab{}.
\newblock \showarticletitle{Foreshadow: Extracting the Keys to the {Intel SGX}
  Kingdom with Transient Out-of-Order Execution}. In
  \bibinfo{booktitle}{\emph{Proceedings of the 27th {USENIX} Security
  Symposium}} \emph{(\bibinfo{series}{{USENIX} Security'18})}.
  \bibinfo{publisher}{{USENIX} Association}.
\newblock


\bibitem[\protect\citeauthoryear{van Schaik, Milburn, \"{O}sterlund, Frigo,
  Maisuradze, Razavi, Bos, and Giuffrida}{van Schaik et~al\mbox{.}}{2019}]%
        {ridl2019}
\bibfield{author}{\bibinfo{person}{Stephan van Schaik}, \bibinfo{person}{Alyssa
  Milburn}, \bibinfo{person}{Sebastian \"{O}sterlund}, \bibinfo{person}{Pietro
  Frigo}, \bibinfo{person}{Giorgi Maisuradze}, \bibinfo{person}{Kaveh Razavi},
  \bibinfo{person}{Herbert Bos}, {and} \bibinfo{person}{Cristiano Giuffrida}.}
  \bibinfo{year}{2019}\natexlab{}.
\newblock \showarticletitle{{RIDL}: Rogue In-flight Data Load}. In
  \bibinfo{booktitle}{\emph{Proceedings of the 40th IEEE Symposium on Security
  and Privacy}} \emph{(\bibinfo{series}{S\&P'19})}. \bibinfo{publisher}{IEEE}.
\newblock


\bibitem[\protect\citeauthoryear{Vila, Abel, Guarnieri, K{\"o}pf, and
  Reineke}{Vila et~al\mbox{.}}{2020}]%
        {flushgeist}
\bibfield{author}{\bibinfo{person}{Pepe Vila}, \bibinfo{person}{Andreas Abel},
  \bibinfo{person}{Marco Guarnieri}, \bibinfo{person}{Boris K{\"o}pf}, {and}
  \bibinfo{person}{Jan Reineke}.} \bibinfo{year}{2020}\natexlab{}.
\newblock \showarticletitle{Flushgeist: Cache Leaks from Beyond the Flush}.
\newblock \bibinfo{journal}{\emph{CoRR}}  \bibinfo{volume}{abs/2005.13853}
  (\bibinfo{year}{2020}).
\newblock


\bibitem[\protect\citeauthoryear{Weisse, Neal, Loughlin, Wenisch, and
  Kasikci}{Weisse et~al\mbox{.}}{2019}]%
        {nda2019weisse}
\bibfield{author}{\bibinfo{person}{Ofir Weisse}, \bibinfo{person}{Ian Neal},
  \bibinfo{person}{Kevin Loughlin}, \bibinfo{person}{Thomas~F. Wenisch}, {and}
  \bibinfo{person}{Baris Kasikci}.} \bibinfo{year}{2019}\natexlab{}.
\newblock \showarticletitle{{NDA}: Preventing Speculative Execution Attacks at
  Their Source}. In \bibinfo{booktitle}{\emph{Proceedings of the 52nd Annual
  IEEE/ACM International Symposium on Microarchitecture}}
  \emph{(\bibinfo{series}{MICRO-52})}. \bibinfo{publisher}{IEEE/ACM}.
\newblock


\bibitem[\protect\citeauthoryear{Yu, Yan, Khyzha, Morrison, Torrellas, and
  Fletcher}{Yu et~al\mbox{.}}{2019}]%
        {STT2019}
\bibfield{author}{\bibinfo{person}{Jiyong Yu}, \bibinfo{person}{Mengjia Yan},
  \bibinfo{person}{Artem Khyzha}, \bibinfo{person}{Adam Morrison},
  \bibinfo{person}{Josep Torrellas}, {and} \bibinfo{person}{Christopher~W.
  Fletcher}.} \bibinfo{year}{2019}\natexlab{}.
\newblock \showarticletitle{{Speculative Taint Tracking (STT): A Comprehensive
  Protection for Speculatively Accessed Data}}. In
  \bibinfo{booktitle}{\emph{{Proceedings of the 52nd Annual IEEE/ACM
  International Symposium on Microarchitecture}}}
  \emph{(\bibinfo{series}{MICRO-52})}. \bibinfo{publisher}{IEEE/ACM}.
\newblock


\end{thebibliography}
